\documentclass[conference]{IEEEtran}
\IEEEoverridecommandlockouts
\usepackage{cite}
\usepackage{amsmath,amssymb,amsfonts}
\usepackage{algorithmic}
\usepackage{graphicx}
\usepackage{stfloats}
\usepackage{textcomp}
\usepackage{xcolor}
\usepackage{tabularx}
\usepackage{array}
\usepackage{url}
\usepackage{listings}
\usepackage{multirow} 

\usepackage[pscoord]{eso-pic}
\newcommand{\placetextbox}[3]{
\setbox0=\hbox{#3}
\AddToShipoutPictureFG*{ \put(\LenToUnit{#1\paperwidth},\LenToUnit{#2\paperheight}){\vtop{{\null}\makebox[0pt][c]{#3}}}
}
}
\placetextbox{.21}{0.055}{\small{978-1-6654-2744-9/21/\$31.00~\copyright2021 IEEE}}

\def\BibTeX{{\rm B\kern-.05em{\sc i\kern-.025em b}\kern-.08em
    T\kern-.1667em\lower.7ex\hbox{E}\kern-.125emX}}
\begin{document}

\title{interID - An Ecosystem-agnostic Verifier Application for Self-sovereign Identity\\

\thanks{}
}


\author{\IEEEauthorblockN{Hakan Yildiz}
\IEEEauthorblockA{\textit{Service-centric Networking} \\
\textit{Technische Universität Berlin}\\
Berlin, Germany \\
hakan.yildiz@tu-berlin.de}
\and

\IEEEauthorblockN{\hspace{1.8cm}Axel Küpper}
\IEEEauthorblockA{\textit{\hspace{1.8cm}Service-centric Networking} \\
\textit{\hspace{1.8cm}Technische Universität Berlin}\\
\hspace{1.8cm}Berlin, Germany \\
\hspace{1.8cm}axel.kuepper@tu-berlin.de}
}

\maketitle

\IEEEpubidadjcol

\begin{abstract}
Self-Sovereign Identity is a transformative paradigm in digital identity management, empowering individuals with full control over their credentials. However, the coexistence of diverse SSI ecosystems, such as the European Digital Identity and the European Blockchain Services Infrastructure, poses significant challenges for cross-ecosystem interoperability due to technological and trust framework differences. This paper introduces \textit{interID}, a modular credential verification application that addresses this fragmentation by orchestrating ecosystem-specific verifier services. Our key contributions include: (1) an ecosystem-agnostic orchestration layer that interfaces with multiple SSI verification services, (2) a unified API that abstracts underlying protocol complexities for service providers, and (3) a practical implementation that bridges three major SSI ecosystems: Hyperledger Indy/Aries, EBSI, and EUDI. Evaluation results demonstrate that interID successfully verifies credentials across all tested wallets with minimal performance overhead, while maintaining a flexible architecture that can be extended to accept credentials from additional SSI ecosystems. This work offers both a technical solution and architectural pattern for achieving interoperability in SSI verifier implementations.
\end{abstract}

\begin{IEEEkeywords}
SSI, Identity, DDI, Interoperability
\end{IEEEkeywords}

\section{Introduction}

Self-Sovereign Identity (SSI) represents a fundamental paradigm shift in digital identities, enabling individuals to independently manage and control their identities without relying on centralized authorities. SSI addresses the critical limitations of traditional digital identity paradigms, such as restricted credential reusability, data privacy concerns, and limited user autonomy.

The promises of SSI have led to the emergence of various technical frameworks, including the Hyperledger Indy\footnote{Hyperledger Indy: \url{https://lf-hyperledger.atlassian.net/wiki/spaces/indy/overview}}/Aries \footnote{Hyperledger Aries: \url{https://lf-hyperledger.atlassian.net/wiki/spaces/ARIES/overview}}, the European Blockchain Services Infrastructure (EBSI)\footnote{European Blockchain Services Infrastructure: \url{https://ec.europa.eu/digital-building-blocks/sites/display/EBSI/Home}}, and the EU Digital Identity (EUDI) framework\footnote{European Digital Identity: \url{https://commission.europa.eu/strategy-and-policy/priorities-2019-2024/europe-fit-digital-age/european-digital-identity_en}}. These frameworks differ significantly in their technological foundations, exchange protocols, credential formats, and trust models \cite{hy_interop_2022}, creating distinct SSI ecosystems that are not interoperable with each other. This fragmentation presents a critical challenge: credentials issued within one SSI ecosystem cannot be validated across ecosystem boundaries, forcing organizations to implement multiple verification systems with increased complexity and costs.

This paper introduces interID, a novel ecosystem-agnostic verifier application designed to bridge these interoperability gaps. interID enables verification of credentials across heterogeneous SSI ecosystems through a unified verification interface, allowing service providers (SPs) to integrate a single solution rather than integrating multiple ecosystem-specific verifier applications. Our primary contributions include:

\begin{itemize}
\item A technology-agnostic credential verification orchestration layer that interfaces with multiple SSI ecosystem-specific verifier services,
\item A unified northbound API that abstracts underlying protocol complexities for SPs,
\item A modular architecture that allows efficient incorporation of additional SSI frameworks without modifying existing components,
\item Performance benchmarks demonstrating the feasibility of adding an orchestration layer for cross-ecosystem credential verification purposes.
\end{itemize}

The rest of the paper is structured as follows. Section \ref{background} provides background on SSI concepts and frameworks. Section \ref{problem} outlines interoperability challenges. Section \ref{related} reviews related work. Sections \ref{design} and \ref{implementation} detail interID's design and implementation. Section \ref{evaluation} presents evaluation results and Sections \ref{future} and \ref{conclusion} discuss future work and conclusions.

\section{Background} \label{background}

SSI marks a significant departure from traditional digital identity paradigms, such as centralized and federated models. Traditional paradigms rely on Identity Providers (IdPs) to manage and verify user identities for other services (relying parties), typically after a login process like username and password. In contrast, SSI empowers individuals by giving them direct control over their own digital identity. This user-centric approach naturally leads to new roles and different ways for organizations and individuals to interact when proving identity.

\subsection{SSI Roles and Interactions}

In the SSI model, identity providers evolve into credential issuers who create and cryptographically sign Verifiable Credentials (VCs). These VCs are issued to identity holders (typically end users). Identity holders store VCs in a wallet application within their domain, e.g., on their smartphones. Identity holders present them to verifiers (SPs or third-party services trusted by SPs) as Verifiable Presentations (VPs). VPs contain critical assurances, including proof of credential ownership and non-revocation, and incorporate features like audience restriction and nonce mechanisms to mitigate replay attacks. Furthermore, verification relies on trust registries, also known as verifiable data registries, that contain the relevant data needed to validate the integrity and authenticity of credentials\cite{preukschat2021self}, without requiring direct communication between the issuers and the verifiers.

\begin{figure}[h!]
\centering
\includegraphics[width=0.9\linewidth]{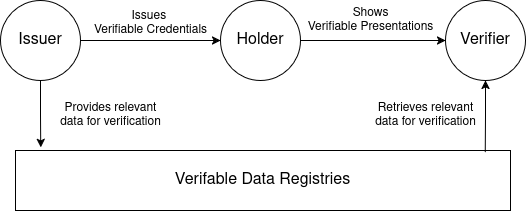}
\caption{High-level SSI model}
\label{fig:ssi_roles}
\end{figure}

Figure \ref{fig:ssi_roles} depicts the interactions between issuers, identity holders, and verifiers as well as the role of verifiable data registries.

\subsection{SSI Frameworks}

The SSI model has been implemented through several technological frameworks, each embodying the core principles of SSI while introducing distinct architectural approaches, protocol choices for exchanging credentials, credential formats, and trust models. These technological divergences have led to multiple siloed ecosystems that, while functionally similar in adhering to the SSI paradigm, differ significantly in their technical underpinnings. In the following, we describe the technological differences of three prominent frameworks.

\subsubsection{Hyperledger Indy/Aries}
Hyperledger Indy provides a purpose-built distributed ledger as a trust registry to store required data for the credential verification, such as identifiers for issuers, credential definitions, and revocation registries \cite{shcherbakov2024hyperledger}. Complementing this, Hyperledger Aries serves as an agent framework facilitating peer-to-peer communication through DIDComm, a secure, encrypted protocol that enables direct interactions between parties without intermediaries \cite{aries2019didcomm}. For credential exchange, this framework employs the Aries Issue Credential protocol for issuance and the Aries Present Proof protocol for verification, both standardized through DIDComm messaging with versions supporting different credential formats \cite{issueCredv1, issuecred2, presentproofv1, presentproof2}. Regarding credential formats, Indy/Aries primarily supports AnonCreds, designed for privacy-preserving use cases with zero-knowledge proofs enabling selective disclosure \cite{anonCreds}, and has added support for W3C JSON-LD VCs with integrity proofs, which provide semantic context through shared vocabularies \cite{ldcredattach}. The Hyperledger Indy/Aries framework underpins national SSI deployments in Canada and Bhutan.

\subsubsection{European Blockchain Services Infrastructure (EBSI)}
EBSI is a collaborative initiative of the European Commission and the European Blockchain Partnership that establishes a blockchain-based infrastructure for delivering cross-border public services throughout the European Union.

EBSI utilizes a permissioned blockchain network operated by EU member states as its trust registry, maintaining records of trusted issuers, schemas, and accreditation information \cite{tan2023verification}. Unlike Indy/Aries, EBSI employs client-server communications using standard web protocols rather than peer-to-peer agent communication. For credential exchanges, EBSI framework contains OpenID for Verifiable Credential Issuance (OID4VCI) for issuance and OpenID for Verifiable Presentations (OID4VP) for verification, extending OpenID Connect standards \cite{oidc4vci, oidc4vp}. The EBSI ecosystem supports W3C JSON-LD VCs with integrity proofs and Selective Disclosure JWT (SD-JWT), which enables presentation of a subset of claims within the VC while maintaining its integrity \cite{sdjwt}.

\subsubsection{EU Digital Identity (EUDI)}
EUDI represents a comprehensive framework aiming to standardize the identification and authentication of natural and legal persons across the member states.

EUDI relies on the existing EU trust infrastructure and trusted lists rather than blockchain technology for its trust registry, maintaining a centralized Trusted Lists Registrar\cite{eudi_arf_1.6}. Similar to EBSI, EUDI employs client-server interactions using standard web protocols for communication. Its exchange protocols include OID4VCI for issuance, and both ISO/IEC 18013-5 for proximity-based exchanges and OID4VP for remote credential presentations \cite{ISO_18013}. EUDI primarily supports Mobile Driver's Licenses (mDoc) based on ISO/IEC 18013-5, which defines standardized data elements and security features \cite{ISO_18013}, and SD-JWTs as credential formats.

\subsection{Framework Comparison}

\begin{table}[h!]
\centering
\caption{Comparison of SSI technology stacks}
\label{table:ssi_comparison}
\begin{tabular}{|p{1.8cm}|p{1.5cm}|p{1.5cm}|p{1.5cm}|}
\hline
\textbf{Features} & \textbf{Indy/Aries} & \textbf{EBSI} & \textbf{EUDI} \\ 
\hline
Credential Formats & AnonCreds, JSON-LD & JSON-LD, SD-JWT & mDL, SD-JWT \\ 
\hline
Exchange Protocols & DIDComm (Present Proof v1/v2) & OID4VCI, OID4VP & OID4VCI, OID4VP, ISO 18013-5 \\ 
\hline
Communication & Peer-to-peer & Client-server & Client-server \\ 
\hline
Trust Registry & Distributed Ledger & Distributed Ledger & PKI-based Trust Registries \\ 
\hline
\end{tabular}
\end{table}

These technological differences create significant challenges for verifiers, who must integrate multiple framework-specific verification services to validate credentials from different ecosystems. The verifier role is particularly critical, as it often serves as the gateway to services, directly interacting with identity holders seeking access to resources.

\section{Problem Statement} \label{problem}

The diversity of SSI frameworks described in Section \ref{background} has led to fragmented ecosystems that cannot easily interoperate despite implementing the same fundamental SSI model. This interoperability challenge manifests itself at multiple technical levels.

\subsection{Interoperability Challenges}

The peer-to-peer DIDComm-based protocols used in Hyperledger Aries are fundamentally different from the client-server OpenID-based protocols used in EBSI and EUDI, making direct communication between wallets and verifiers from different ecosystems impossible. Different credential formats (AnonCreds, JSON-LD, mDoc) have distinct data structures and verification mechanisms, while divergent trust models (blockchain-based vs. PKI-based) create incompatible trust validation paths.

These technical differences prevent the use of credentials across the ecosystem. Consider a real-world scenario: a Canadian citizen with an Indy/Aries-based eID attempts to authenticate to a European service that only supports credentials from EUDI ecosystem. Despite possessing a valid government-issued credential, The authentication fails because the European service's verifier can neither request the credential nor process the Canadian credential's format or its root of trust.

\subsection{Organizational and User Challenges}

For organizations deploying SSI verification capabilities, the current fragmentation creates substantial challenges. First, SPs must implement and maintain multiple verification systems, which increases integration costs. Moreover, each new verification system demands specialized knowledge, increasing the required human resources. Finally, users experience inconsistent verification flows depending on the underlying framework of their credentials, leading to confusion and a reduction in user acceptance.

\subsection{Need for an Interoperable Verification Solution}

A unified verifier solution that can validate credentials across multiple SSI ecosystems would address these challenges. Such a solution would abstract the complexities of different verification protocols and credential formats, present a consistent interface to SPs, support multiple trust models simultaneously, and scale efficiently to incorporate additional frameworks as the SSI landscape evolves.

The interID application presented in this paper directly addresses this need by providing a multi-ecosystem credential verification orchestration layer that enables SPs to integrate a single solution rather than multiple ecosystem-specific verifier services.

\section{Related Work} \label{related}

Interoperability in SSI ecosystems has been addressed through both standardization efforts and academic research.

\subsection{Standardization Efforts}

Standardization initiatives have primarily focused on achieving interoperability by creating profiles. A profile typically refers to a selected combination of protocols, credential formats, and trust anchors, along with a tighter configuration of those protocols. This tighter configuration is necessary because many protocols in the SSI space contain numerous optional parameters or allow for multiple configuration options, which hinders interoperability without explicit alignment.

The Hyperledger community developed the Aries Interoperability Profiles (AIP) \cite{aip}, which define standardized agent interactions within the Hyperledger Aries ecosystem but do not address the verification of credentials from non-Aries ecosystems.

The Decentralized Identity Foundation (DIF) introduced the Wallet and Credential Interaction (WACI) profile \cite{waci} and Presentation Exchange specifications. Although WACI is designed to be credential-agnostic, implementations typically remain bound to specific ecosystems that rely on peer to peer communications.

OpenID Connect-based standards, including OID4VCI and OID4VP \cite{oidc4vci, oidc4vp}, have gained widespread adoption in client-server SSI models. The High-Assurance Interoperability Profile (HAIP) \cite{haip} for EUDI and the EBSI Interoperability Profile (EBSI v3)\cite{EBSIInteropProfile} enable credential issuance and verification capabilities within their contained ecosystems.

\subsection{Academic Research}

Academic research on SSI interoperability has explored multiple dimensions, offering insights into the complex challenges of achieving seamless credential verification across different technological ecosystems.

From a conceptual perspective, Yildiz et al. developed a comprehensive reference model for SSI interoperability \cite{yildiz2023interoperable, hy_interop_2022}, identifying critical requirements such as protocol alignment, trust establishment, and secure communications. However, their work primarily focused on theoretical frameworks, falling short of providing concrete implementation strategies for cross-ecosystem verifiers.

Moreover, from a comparative studies standpoint, researchers have systematically analyzed the technological landscape. Pava-Díaz et al. conducted an extensive analysis of nine distributed ledger-based SSI frameworks \cite{pava2024ssi}, revealing significant divergences in verification processes. Building upon this, Čučko and Turkanović's systematic mapping study \cite{cucko2021systematic} critically highlighted that verifier-level integration between ecosystems remains a largely underdeveloped area of academic research.

From an implementation approaches perspective, Grüner et al. provided an in-depth analysis of SSI interoperability \cite{SSIinteropPaperHPI}. Their research meticulously evaluated major frameworks against key criteria including data format compatibility, trust model characteristics, and credential exchange mechanisms. While they compellingly argued that true interoperability necessitates alignment at both technical and conceptual levels, their contribution remained predominantly theoretical, without offering a practical implementation blueprint.

Furthermore, Krul et al. made a significant contribution by identifying divergent trust models as a fundamental barrier to cross-platform verification \cite{krul2024trust}. This insight underscores the complexity of achieving interoperability in decentralized identity systems, highlighting the need for more sophisticated integration approaches.

\subsection{Research Gap}

Our analysis reveals a significant gap between theoretical interoperability models and practical verification implementations. While standardization efforts have made progress within specific ecosystems, and academic research has established conceptual frameworks, there is a critical absence of implemented solutions that enable verification across heterogeneous SSI ecosystems. This absence of practical cross-ecosystem verification mechanisms represents a significant barrier to SSI adoption in multi-framework environments where SPs must navigate technological fragmentation.

\section{Concept and Design} \label{design}

To address this research gap, we propose interID, an orchestration layer that enables verification of credentials across heterogeneous SSI ecosystems through architectural abstraction rather than protocol harmonization. The fundamental insight driving interID's design is that ecosystem interoperability can be achieved at the orchestration layer without requiring modifications to existing credential exchange protocols, credential formats, or trust models. By integrating with framework-specific verification services and exposing a unified northbound interface, interID creates a bridge between fragmented ecosystems from the verifier perspective and allows SPs to integrate with their backend conveniently.

\subsection{Design Requirements}

The design of interID is guided by four essential non-functional requirements:

\textbf{Interoperability:} interID must support multiple SSI ecosystems simultaneously, enabling the verification of credentials across technological and ecosystem boundaries. This can be achieved through ecosystem-specific adapters that translate between interID's unified API and the distinct request parameters and bodies of each verification service endpoints.


\textbf{Reliability:} The system must reliably transmit verification results to SPs by consistently delivering accurate and cryptographically verifiable validation results that encompass the requested attributes. This reliability requirement is essential, as SPs base authorization decisions for access to goods and services on these verification results, making the trustworthiness of interID's output a critical operational parameter.

\textbf{Extensibility:} The overall system must be modular. The modular architecture enables straightforward integration of additional SSI frameworks without disrupting existing functionality. New verification services can be added by the implementation of well-defined interfaces, allowing interID to evolve alongside the SSI landscape.

\textbf{Performance:} As a critical component in authentication flows, the system must only add minimal overhead beyond what is required by the underlying verification services themselves.

\subsection{System Architecture}

interID employs a three-layered architecture designed to abstract the complexities of different SSI ecosystems:

\begin{itemize}
\item \textbf{Service Layer:} Integrates with framework-specific verifier services responsible for validating credentials according to their respective protocols, formats, and trust models.

\item \textbf{Controller Layer:} Orchestrates the verification process across different services, maps unified proof templates to framework-specific formats, and provides consistent response handling regardless of the underlying verification service used.

\item \textbf{Presentation Layer:} Offers a user interface for administrators to configure proof templates and for identity holders to initiate verification processes with their preferred wallet application.
\end{itemize}

This layered approach enables interID to maintain a clean separation of concerns while providing extensibility for future SSI frameworks and verification services.

\begin{figure}[ht]
\centering
\includegraphics[width=0.85\linewidth]{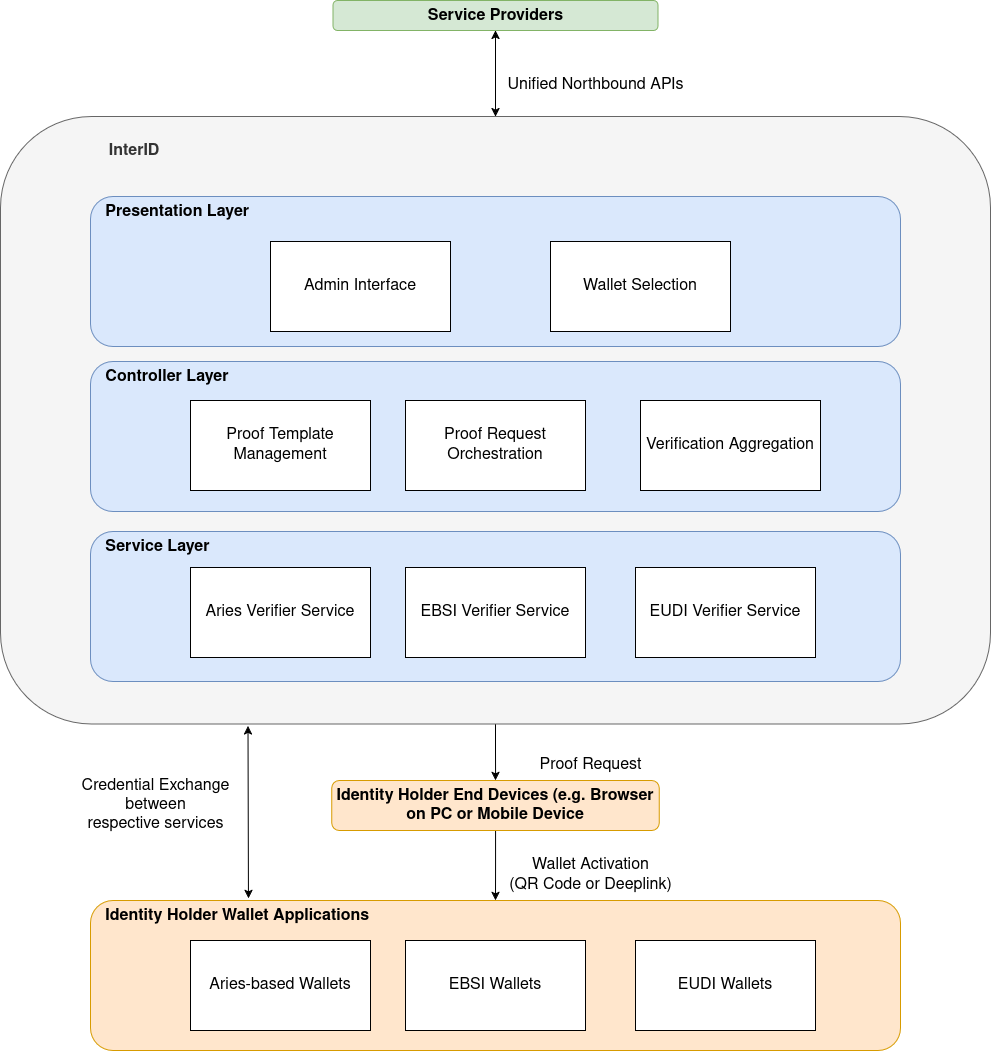}
\caption{interID high-level architecture showing orchestration of multiple verification services}
\label{fig:interID_architecture}
\end{figure}

Figure \ref{fig:interID_architecture} illustrates the high-level architecture of interID, showing how it bridges multiple SSI ecosystems through their respective verification services.

\subsection{Core Capabilities}

\subsubsection{Proof Template Management}

interID introduces the concept of proof templates that allow SPs to define verification requirements once and apply them across multiple SSI ecosystems. A proof template serves as a container that can include contents of the verification request for each supported ecosystem.

\begin{itemize}
\item For Hyperledger Aries: AnonCreds proof requests via Aries Present Proof protocol v1 or v2 and JSON-LD presentation requests using Aries Present Proof protocol v2
\item For EBSI: JSON-LD or SD-JWT presentation requests using OID4VP in EBSI profile configuration
\item For EUDI: mDoc or SD-JWT presentation requests using OID4VP in HAIP configuration
\end{itemize}

This approach enables SPs to maintain consistent verification requirements across different frameworks without managing multiple verification systems.

\subsubsection{Credential Exchange Workflow}

The credential exchange process follows a consistent pattern regardless of the underlying SSI ecosystem:

\begin{enumerate}
\item A SP initiates a proof request through interID's API, referencing a previously created proof template
\item The identity holder selects a compatible wallet application from the options presented by interID
\item interID invokes the appropriate verification service based on the selected wallet type with the proof request stored in the template
\item The verification service and the identity holder conduct the credential exchange using the protocol and format supported by the selected wallet
\item The verification service validates the presented credentials according to ecosystem-specific rules
\item interID standardizes the verification results and returns them to the SP in a consistent and verifiable format
\end{enumerate}

\section{Implementation} \label{implementation}

interID integrates multiple verifier services: specifically Hyperledger Aries Cloud Agent Python (ACA-Py), Walt.id, and the EUDI Reference Implementation, ensuring broad compatibility with with Hyperledger-based, EUDI, and EBSI ecosystems. Each verifier service is deployed as a Docker container to provide a consistent and straightforward deployment across different environments. Furthermore, each verifier service supports credential resolution mechanisms suitable for their targeted SSI ecosystem.

\subsection{Verifier Backends}

InterID integrates three specialized verifier services as Docker containers, each targeting specific SSI ecosystems. ACA-Py supports Hyperledger Indy/Aries credentials through both Present Proof v1 (for AnonCreds) and v2 protocols (for AnonCreds and JSON-LD), leveraging multi-ledger configuration to interact with various trust registries. Walt.id provides verification capabilities for the EBSI ecosystem using OID4VP, supporting both JSON-LD and SD-JWT credential formats in EBSI v3 profile configuration. The EUDI verifier handles mDoc and SD-JWT credentials according to the HAIP. Each verifier service implements credential resolution mechanisms appropriate for its ecosystem's trust model.






\subsection{interID Backend}

The backend implements a modular, service-oriented architecture using Node.js with TypeScript, comprising four core modules.

This modular approach enables clean separation of concerns while facilitating extension with additional SSI ecosystems.


\textbf{Components} module encapsulates domain-specific logic, including handling proof requests and managing proof templates. It defines the controllers and endpoints responsible for processing incoming requests, orchestrating interactions, and managing business logic.

\textbf{Routes} define RESTful API endpoints with standardized request/response patterns, facilitating consistent interactions with frontend applications and service providers. 

\textbf{Services} module manages integration logic with the containerized verification services. This module also responsible for handling the credential verification processes and aggregates verification results.

\textbf{Utils} provide functions for cross-cutting concerns, including structured logging via Winston, Redis-based session management for correlation between verification requests and responses, and input validation through Zod schema validation.


\subsection{Template Creation and Proof Request Handling}
interID simplifies complex and proof request processes through robust template management and streamlined proof request handling. These capabilities enhance consistency, reduce integration complexity, and facilitate interoperability across multiple SSI frameworks. This subsystem comprises three primary elements: Proof Template Creation, Proof Request Templates, and Proof Request and Response handling.

\subsection{Proof Templates and Proof Request Handling}

InterID implements proof templates as reusable configurations that define verification requirements across multiple SSI ecosystems. Templates persist in MongoDB using a flexible schema that allows ecosystem-specific sections for each supported verification protocol. The ProofTemplateSchema defines a unified container with fields for different protocol implementations: v1 and v2 for AnonCreds via Aries Present Proof protocols, dif for JSON-LD credentials, waltid for EBSI-compliant verification, and pid for EUDI credentials. This design enables service providers to define verification requirements once while applying them across heterogeneous ecosystems.

The implementation maps unified attribute requests to ecosystem-specific formats through specialized query builders. \texttt{AnonCredQueryBuilder} transforms generic attribute requests into AnonCreds proof requests with requested\_attributes and requested\_predicates sections, handling comparative operators through appropriate predicate mapping. \texttt{DIFQueryBuilder} generates DIF Presentation Exchange structures with appropriate constraints and input descriptors for JSON-LD verification. \texttt{WaltQueryBuilder} addresses the specific requirements of the EBSI implementation in Walt.id, while \texttt{PIDQueryBuilder} generates format-specific requests for both mDoc and SD-JWT within the EUDI framework. These transformations abstract the protocol-specific complexities from service providers while enabling verification across technological boundaries.

Proof request handling implements distinct verification flows for each ecosystem. Aries verification involves a two-phase process of creating a presentation request and generating an out-of-band invitation with the request attached. Walt.id verification posts requests to its verification endpoint with webhook callbacks for results, while EUDI verification constructs appropriate presentation definitions for both mDoc and SD-JWT formats. Session management utilizes Redis with a 7200-second expiration to maintain stateful connections between initial requests and eventual verification results, enabling the system to correlate asynchronous verification outcomes with specific user sessions.

\subsubsection{Proof Template Creation}

Proof templates enable SPs to create standardized and reusable structures for initiating credential verification requests. Each proof template is uniquely identified by a universally unique identifier (UUID), acting as a container for proof request queries specific to each integrated verifier service. The implementation utilizes MongoDB with Mongoose as the object-document mapper, persisting templates with the following schema structure:

\begin{lstlisting}[basicstyle=\footnotesize\ttfamily, breaklines=true, caption={Proof Template Schema Definition}, label=lst:proof_template_schema]
const ProofTemplateSchema = new Schema({
    _id: {
        type: String,
        required: true,
        default: () => uuidv4()
    },
    name: {
        type: String,
        required: true,
    },
    createdAt: {
        type: Date,
        default: Date.now,
    },
    v1: { type: Schema.Types.Mixed, required: false },
    v2: { type: Schema.Types.Mixed, required: false },
    dif: { type: Schema.Types.Mixed, required: false },
    waltid: { type: Schema.Types.Mixed, required: false },
    pid: { type: Schema.Types.Mixed, required: false }
});
\end{lstlisting}

This schema design allows a single template to contain verification configurations for all supported SSI ecosystems, with each field targeting a specific protocol and, in the case of ACA-Py, credential format. The backend maintains these templates, facilitating simple retrieval, modification, and reuse, which significantly reduces repetitive configuration tasks.

\subsubsection{Proof Request Template}

Proof request templates encapsulate detailed queries that specify the credentials and claims requested by the verifier from identity holders. Due to varying syntax requirements among verifier services (ACA-Py, Walt.id, EUDI), interID abstracts these complexities by offering a unified API. Internally, the backend service dynamically maps unified templates to backend-specific proof request structures:

\begin{itemize}
\item \textbf{ACA-Py:} Supports proof request language for AnonCreds (v1 and v2 schemas, allocated for Present Proof Protocols v1 and v2, respectively). The implementation uses the \texttt{AnonCredQueryBuilder} to transform generic attribute requests into structures with \texttt{requested\_attributes} and \texttt{requested\_predicates}, supporting complex operators like greater-than or less-than comparisons. ACA-Py also supports DIF Presentation Exchange for JSON-LD credentials (dif schema, allocated for Present Proof v2), whereas the \texttt{DIFQueryBuilder} transforms the attribute requests to structures supported by the ACA-Py.

\item \textbf{Walt.id:} Utilizes DIF Presentation Exchange query language via the OID4VP protocol, in accordance with the EBSI profile. However, walt.id request parameters and body is different than what is expected from DIF Presentation Exchange. Therefore, the \texttt{WaltQueryBuilder} maps the presentation request query to the format processable by walt.id verifier backend.

\item \textbf{EUDI Reference Implementation:} Employs OID4VP protocol with DIF Presentation Exchange definitions aligned with the HAIP. The \texttt{PIDQueryBuilder} generates format-specific requests for both mDoc and SD-JWT credential formats with appropriate descriptors and constraints.
\end{itemize}

\begin{figure}[ht]
\centering
\includegraphics[width=0.85\linewidth]{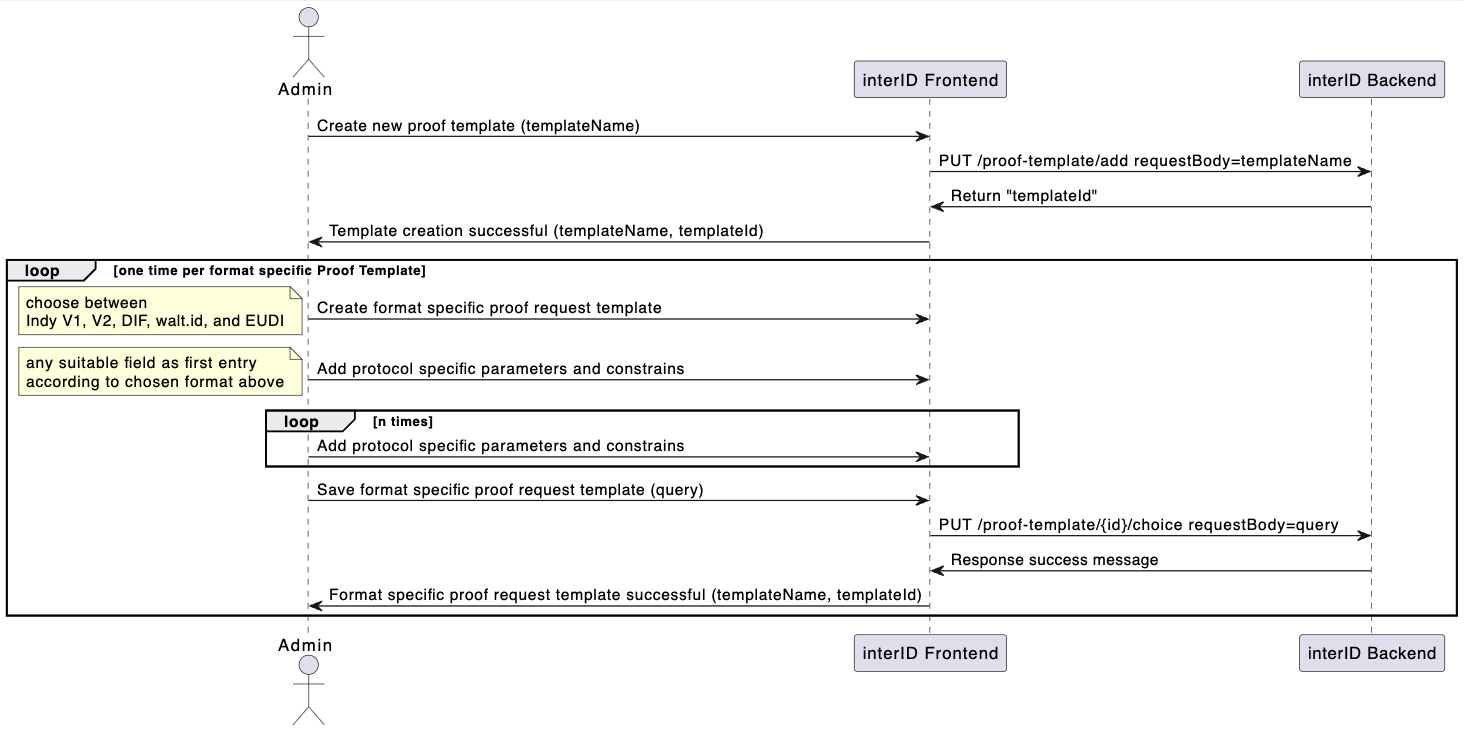}
\caption{Sequence diagram of credential verification through interID}
\label{fig:proof_template}
\end{figure}

Figure \ref{fig:proof_template} illustrates how proof templates and proof request templates are created.

\subsubsection{Proof Request and Response Handling}

To initiate proof verification, SPs invoke a standardized endpoint provided by interID, referencing a pre-defined proof template. The \texttt{ProofRequestController} orchestrates the verification process through a unified RESTful API while accommodating the distinct protocols of each ecosystem:

\begin{itemize}
\item For Hyperledger Aries, verification involves a two-phase process:
  \begin{itemize}
    \item Creating a presentation request via ACA-Py's \texttt{/present-proof/create-request} endpoint
    \item Generating an out-of-band invitation via \texttt{/out-of-band/create-invitation} endpoint with the presentation request attached.
  \end{itemize}

\item For the Walt.id EBSI implementation:
  \begin{itemize}
    \item The controller posts the verification request to the \texttt{/openid4vc/verify} endpoint
    \item It captures the state identifier from the response URL for session tracking
  \end{itemize}

\item For EUDI verification:
  \begin{itemize}
    \item The controller constructs a PID-specific verification request to the \texttt{/ui/presentations} endpoint using appropriate presentation definition parameters
    \item It handles format-specific verification paths for both mDOC and SD-JWT credential types
  \end{itemize}
\end{itemize}

Upon generating verification requests for each ecosystem, interID supports two distinct verification flows that accommodate varying user device configurations. In the \textbf{same device flow}, when verification is initiated on the device hosting the wallet, interID renders ecosystem-specific deeplinks such as \texttt{eudi-openid4vp://} or \texttt{didcomm://} that directly launch the compatible wallet application.

Conversely, in scenarios where the wallet resides on a different device from the verification initiation point, \textbf{cross-device flow} is supported by generating QR codes that encode the identical proof request, enabling identity holders to complete the verification process by scanning the code with their wallet-equipped device.

All verification processes utilize Redis for session management, maintaining stateful connections between initial request, user interaction, and verification completion. Redis stores mapping between session identifiers, verification exchanges, and user login information, with appropriate timeout settings to prevent resource leaks.

Lastly, upon successful verification, interID aggregates and returns structured responses detailing validation outcomes in a JWT. This structured response provides clear, actionable feedback for SPs, facilitating informed decision-making processes.

\begin{figure*}[ht]
\centering
\includegraphics[width=0.85\linewidth]{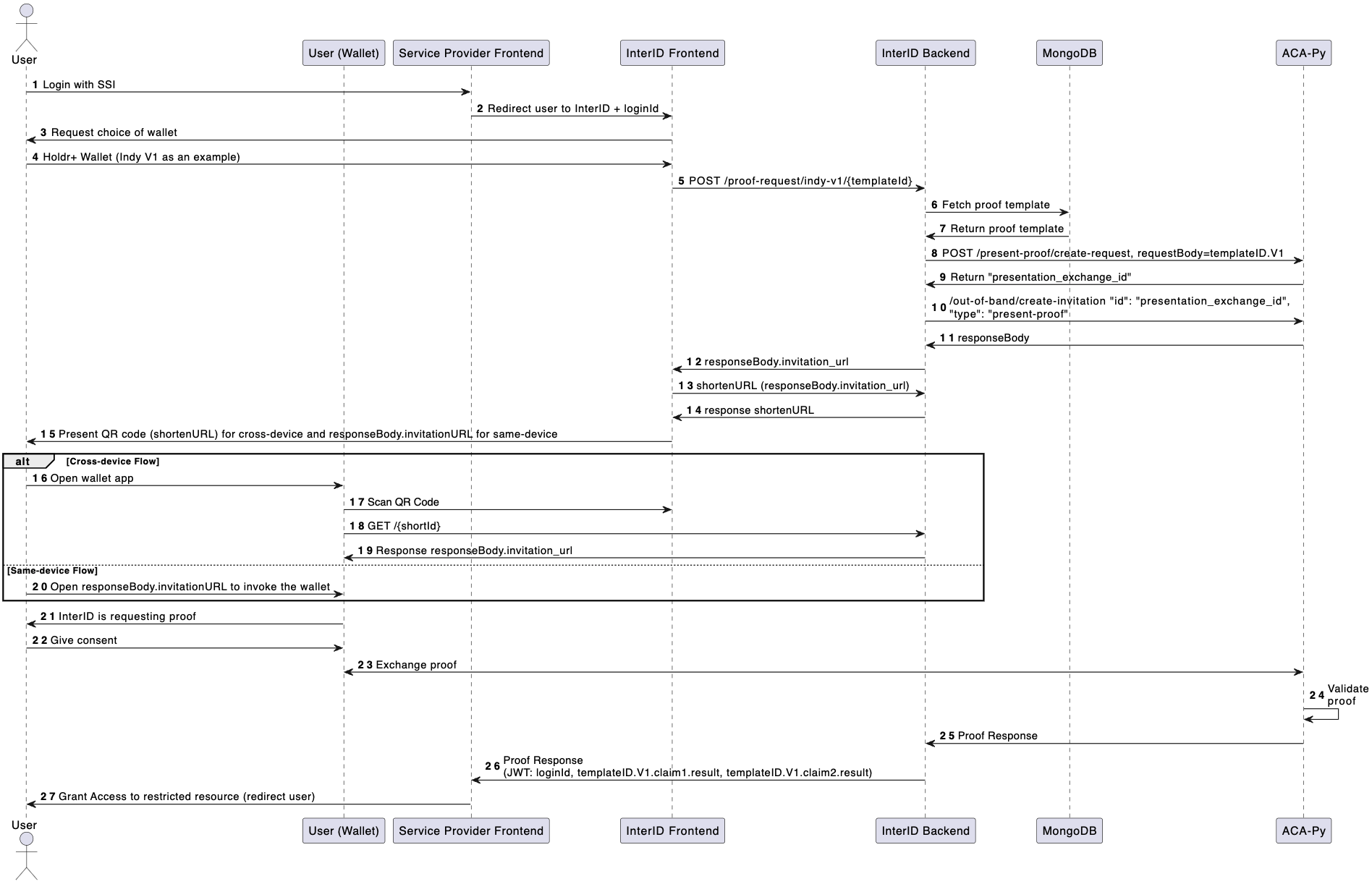}
\caption{Sequence diagram of credential verification through interID}
\label{fig:verification_flow}
\end{figure*}

Figure \ref{fig:verification_flow} illustrates an examplary sequence, highlighting how interID abstracts the ecosystem-specific details from both the SP and the identity holder.

\subsection{Frontend Implementation}
The frontend implementation serves as a multi-purpose graphical user interface (GUI) to support various interID workflows. Its primary functions include enabling SPs to create and manage proof templates, facilitating the configuration of service-specific proof request templates, and providing identity holders with an interface to select their preferred wallet during verification processes.

Moreover, the interface offers granular control over proof request parameters for SPs, allowing them to specify which credential claims require validation, designate acceptable issuers, and define constraint conditions (e.g., equivalence, greater than or equal to, less than or equal to) for specific claim values. This functionality enables SPs to precisely define the verification requirements without requiring them to know protocol-specific query parameters.

The implementation utilizes React with TypeScript for robust frontend development, employing Tailwind CSS for interface styling and Vite as the build optimization tool. The architecture follows a component-based design pattern to enhance maintainability and code reusability. Backend communication is implemented through RESTful API endpoints that handle proof template management, proof request initiation, and verification result processing.

\subsection{Implementation Challenges}
The implementation of interID encountered several challenges that highlight the current state of the SSI ecosystem and provide valuable insights for similar interoperability initiatives.

The proof requests generated by the ACA-Py encountered significant data transmission challenges during interID's implementation. Specifically, the out-of-band invitation, when encoded in QR codes, exceeded processing capabilities of the Aries wallets under examination. To mitigate this technical constraint, interID integrated a URL shortening mechanism that dynamically generates compact endpoint references. This approach enables the backend to invoke and retrieve the full out-of-band invitation, effectively circumventing the data transmission limitations inherent in the current Aries wallet implementations.

Moreover, the continuous evolution of reference implementations, particularly for Walt.id and the EUDI framework, presented ongoing integration challenges. During our implementation period, both frameworks underwent significant updates that altered API behaviors and protocol implementations. These changes required multiple adaptations to our integration code and underscored the importance of developing loosely coupled interfaces that can accommodate evolving specifications without requiring extensive refactoring of the orchestration layer.

\section{Evaluation} \label{evaluation}

To ensure that the implemented system meets its design objectives and operational requirements, interID is evaluated based on four criteria: interoperability, performance, reliability, and extensibility.

\subsection{Interoperability}

To assess interoperability, comprehensive tests were designed to validate credential verification across the supported technology stacks: Hyperledger Aries, Walt.id (EBSI), and the EUDI reference implementation. The evaluation focused on verifying that interID effectively abstracts backend-specific complexities, offering a unified and standardized northbound interface to SPs.

Specifically, interoperability was tested with two ACA-Py-compatible wallets configured for different Indy ledgers as trust anchors: Indicio Test Network and BC Gov Test Network. Credential verification tests successfully requested and validated AnonCreds via the Aries Present Proof v1 Protocol (AIP 1.0). Additionally, another ACA-Py instance was configured to handle JSON-LD Credentials, tested through the Aries Present Proof v2 Protocol (AIP 2.0).

In the context of EBSI, interoperability was validated by testing JSON-LD Credential presentations using the OID4VP protocol, specifically configured to the EBSI v3 Profile. Credential exchanges were successfully conducted using the Walt.id web wallet, demonstrating effective interoperability with this ecosystem.

Lastly, interoperability within the EUDI framework was tested using both iOS and Android versions of the reference implementation wallets. Credential exchanges were verified for Personal Identification (PID) requests, The SD-JWT format of PID could be validated while the mDoc validation fails due to missing...

Results demonstrated successful credential exchanges and validations across all integrated protocols and credential formats, confirming that interID effectively achieves its interoperability objectives.

Table \ref{tab:interoperability_results} summarizes the credential formats and exchange protocols verified by the individual verifier services and illustrates interID's collective capability to validate all credentials:

\begin{table}[!htbp]
\centering
\caption{Interoperability Test Results}
\label{tab:interoperability_results}
\resizebox{\columnwidth}{!}{
\begin{tabular}{|p{4cm}|c|c|c|c|}
\hline
\textbf{Wallet} & \textbf{ACA-Py} & \textbf{Walt.id (EBSI)} & \textbf{EUDI Ref. Impl.} & \textbf{interID} \\
\hline
Indicio Wallet & \checkmark &  &  & \checkmark \\
\hline
BC Gov Wallet & \checkmark &  &  & \checkmark \\
\hline
ACA-Py JSON-LD Wallet & \checkmark &  &  & \checkmark \\
\hline
Walt.id Web Wallet &  & \checkmark &  & \checkmark \\
\hline
EUDI Wallet (iOS/Android) &  &  & \checkmark & \checkmark \\
\hline
\end{tabular}}
\end{table}

\subsection{Performance}
The performance evaluation quantified the overhead introduced by interID compared to direct interactions with individual verifier backends. The benchmarks were designed to determine whether interID's abstraction layer introduces significant latency in credential verification processes.

Since end-to-end credential validation involves user-device interactions and network variability that could obscure interID's direct performance impact, our measurements focused specifically on the initiation of proof requests. Each test was executed 100 times to establish statistically reliable averages and eliminate outliers. The evaluation targeted each endpoint responsible for creating proof request templates and forwarding them to the underlying verifier services.

To quantify interID's overhead, we measured the request and response times when (1) routing through interID and (2) directly accessing the respective verifier service endpoints. All tests were conducted in a controlled local environment using an Intel i5-1135G7 processor with 24 GB of RAM.

Table~\ref{tab:performance_results} presents the comparative latency results, illustrating interID's overhead relative to individual verifier services. The overhead percentage was calculated as the relative increase in processing time when using interID compared to native implementation ((interID time - native time) / native time × 100\%).

\begin{table}[ht]
\centering
\caption{Performance Comparison: interID Overhead vs. Native Implementation (in ms)}
\label{tab:performance_results}
\resizebox{\columnwidth}{!}{
\begin{tabular}{|l|ccc|ccc|ccc|}
\hline
\multirow{2}{*}{\textbf{Verifier}} & \multicolumn{3}{c|}{\textbf{With interID}} & \multicolumn{3}{c|}{\textbf{Native}} & \multicolumn{3}{c|}{\textbf{Overhead (\%)}} \\
\cline{2-10}
 & \textbf{Avg} & \textbf{Min} & \textbf{Max} & \textbf{Avg} & \textbf{Min} & \textbf{Max} & \textbf{Avg} & \textbf{Min} & \textbf{Max} \\
\hline

Aries V1 & 69.44 & 55 & 108 & 59.29 & 44 & 79 & 17.12 & 25.00 & 36.71 \\
Aries V2 & 47.38 & 36 & 117 & 41.92 & 28 & 131 & 13.02 & 28.57 & -10.69 \\
Aries DIF & 49.18 & 36 & 131 & 42.94 & 29 & 127 & 14.53 & 24.14 & 3.15 \\
walt.id & 15.90 & 10 & 28 & 9.84 & 5 & 19 & 61.60 & 100.00 & 47.37 \\
EUDI Verifier & 137.07 & 128 & 175 & 129.46 & 125 & 167 & 5.88 & 2.40 & 4.79 \\
\hline
\end{tabular}}
\end{table}

The results demonstrate that interID introduces varying degrees of overhead across different verifier implementations. While the walt.id integration showed the highest relative overhead (61.60\% on average), it is important to note that the absolute latency remains low (15.90ms with interID versus 9.84ms native). The EUDI Verifier demonstrated the lowest relative overhead (5.88\%), and Aries integration showed moderate overhead ranging from 13.02\% to 17.12\%. These findings indicate that while interID's abstraction layer does introduce some processing overhead, the absolute latency values remain within acceptable ranges for the feasibility of using a verification orchestrator to achieve cross-ecosystem verifier interoperability.

\subsection{Reliability}

Reliability evaluation focused on the consistency and integrity of verification results transmitted to SPs through JWT responses. Our testing methodology compared the attribute claims requested in proof templates against those delivered following successful verification across all supported SSI ecosystems. This assessment confirmed that interID consistently delivers precisely the requested attribute claims without data loss, ensuring that SPs receive accurate information for making authorization decisions.

\subsection{Extensibility}

The extensibility evaluation analyzed how easily new verifier services could be integrated into interID. This capability was primarily assessed by our own integration of multiple verifier services. Specifically, extending interID with additional verifier services can be achieved efficiently due to the clearly structured backend architecture. Integrating a new verifer service typically involves the following steps:

\begin{itemize}
\item \textbf{Adding a New Dockerized Container:} A new verifier service can be encapsulated within its Docker container. Additionally, the docker-compose.yml file has to be configured to start the newly added docker container. 

\item \textbf{Environment Configuration:} The integration of additional services involves updates to environment configuration files (.env). These files define endpoint URLs and other service-specific parameters, enabling simple configuration and management of new verifier services.

\item \textbf{Controller Layer Adjustments:} Within the controller layer, new controllers must be created to handle the interactions with the additional verifier service. Clearly defined interfaces and modularized controllers facilitate minimal and localized changes, ensuring integration without affecting existing functionalities.

\item \textbf{Proof Template and Proof Request Modifications:} Proof template schemas and proof request controllers require minor updates to accommodate any new protocol specifics introduced by the new verifier service. The modular schema and controller structures allow rapid incorporation of such changes without changes in the architecture.

\item \textbf{Frontend Integration:} Minor updates to the frontend are typically required to allow SPs and users to select and interact with the newly added verification service, leveraging the existing modular UI framework and API interfaces.

\end{itemize}

\section{Future Work} \label{future}

Future development efforts will focus on transitioning interID into a Software-as-a-Service (SaaS) solution, removing the necessity for SPs to deploy interID within their own infrastructure. Offering interID as a cloud-based service, complemented by user-friendly integration interfaces and user management capabilities, has the potential to significantly enhance its accessibility and adoption. This transition could further accelerate the broader acceptance and utilization of SSI among SPs.

Additionally, the current implementation transmits proof verification results to SPs as JWT. While sufficient for deployments within a secure and controlled service-provider domain, this method lacks the necessary security guarantees for a SaaS environment. Future implementations should explore and integrate established secure token exchange mechanisms such as Security Assertion Markup Language (SAML) or OpenID Connect (OIDC). Employing these standards would allow SPs to reliably authenticate responses and verify their integrity, analogous to how ID tokens are utilized in OIDC-based workflows. This approach would provide E2E interoperability between the SPs relying on SAML or OIDC for authentication and various SSI ecosystems via interID.

\section{Conclusion} \label{conclusion}

This paper introduced \textit{interID}, a verifier application designed to operate across heterogeneous SSI ecosystems by supporting multiple frameworks in different ecosystem-specific profiles and configurations. By abstracting protocol-specific behaviors and leveraging a layered architecture, interID enables credential verification irrespective of the underlying SSI ecosystem. Our approach addresses a pressing need in the SSI landscape: verifier-side interoperability. Through practical implementation and early integration efforts with European EUDI and EBSI initiatives and Hyperledger Indy based ecosystems, we demonstrate the feasibility and impact of such a bridge.

\section{Acknowledgement} \label{acknowledgement}

\bibliography{literature}

\end{document}